\documentclass[aps,prl,amsmath,amssymb,reprint,twocolumn,nofootinbib,superscriptaddress]{revtex4-2}

\usepackage{graphicx}
\usepackage{bm}
\usepackage{xcolor}
\usepackage{hyperref}
\usepackage{cleveref}
\usepackage{lmodern}
\usepackage{microtype}
\usepackage{enumitem}

\hypersetup{
    colorlinks=true,
    linkcolor=blue,
    citecolor=blue,
    }

\newsavebox{\foobox}
\newcommand{\slantbox}[2][0]{\mbox{%
        \sbox{\foobox}{#2}%
        \hskip\wd\foobox
        \pdfsave
        \pdfsetmatrix{1 0 #1 1}%
        \llap{\usebox{\foobox}}%
        \pdfrestore
}}
\newcommand\unslant[2][-.25]{\slantbox[#1]{$#2$}}

\newcommand{\thetaEff}{\theta_\text{eff}}
\newcommand{\thetaYukawa}{\theta_Y}
\newcommand{\thetaTotal}{\bar\theta}
\newcommand{\thetaStrong}{\theta_{\text{QCD}}}
\newcommand{\dint}{\,\text{d}}
\newcommand{\given}{\,|\,}
\newcommand{\muev}{\,\unslant{\mu}\text{eV}}
\newcommand{\mev}{\,\text{meV}}
\newcommand{\ev}{\,\text{eV}}
\newcommand{\gev}{\,\text{GeV}}
\newcommand{\lag}{\mathcal{L}}
\newcommand{\mpl}{M_\text{Pl}}
\newcommand{\ecm}{\,e\,\text{cm}}
\newcommand{\Nanom}{N_\text{DW}}
\newcommand{\BG}{\Delta}

\renewcommand{\Re}{\operatorname{Re}}
\renewcommand{\Im}{\operatorname{Im}}

\begin{document}


\title{Quality control for axions and ALPs}

\author{Andrew Cheek}
\email{acheek@sjtu.edu.cn}
\affiliation{State Key Laboratory of Dark Matter Physics, Tsung-Dao Lee Institute \& School of Physics and Astronomy, Shanghai Jiao Tong University, Shanghai 200240, China}
\affiliation{Key Laboratory for Particle Astrophysics and Cosmology (MOE) \& Shanghai Key Laboratory for Particle Physics and Cosmology, Shanghai Jiao Tong University, Shanghai 200240, China}

\author{Andrew Fowlie}
\email{andrew.fowlie@xjtlu.edu.cn}
\affiliation{X-HEP Laboratory, Department of Physics, School of Mathematics and Physics, Xi'an Jiaotong-Liverpool University,  111 Ren'ai Road, Suzhou Dushu Lake, Science and Education Innovation District, Suzhou Industrial Park, Suzhou 215123, China}

\author{Gonzalo Herrera}
\email{gonzaloh@mit.edu}
\affiliation{Physics Department and Kavli Institute for Astrophysics and Space Research, Massachusetts Institute of Technology, Cambridge, MA 02139, USA}
\affiliation{Harvard University, Department of Physics and Laboratory for Particle Physics and Cosmology, Cambridge, MA 02138, USA}

\begin{abstract}
Axions and axion-like particles (ALPs) are protected by Peccei-Quinn (PQ) symmetries that quantum gravity is expected to break. Modeling quantum gravity by Planck-suppressed PQ-breaking operators with unsuppressed Wilson coefficients and random phases, we quantify the fine-tuning required for an acceptable strong CP phase or a given ALP mass. For the QCD axion to account for the observed dark matter abundance at $f_a \simeq 10^{11}\gev$, PQ-breaking operators must be absent up to mass dimension $D \gtrsim 12$. We show that the residual strong CP phase could be measurable in upcoming neutron electric dipole moment searches. For ALPs, we map the mass-decay constant plane by the degree of UV protection required, and find that parts of the parameter space targeted by future laboratory experiments are already fine-tuned at the part-per-million level or worse, or equivalently, require PQ quality to be protected up to dimension $D\gg 10$. We argue that quality, not mass alone, is the central naturalness question for the axion program.
\end{abstract}

\maketitle


\section{Introduction}

The QCD axion~\cite{Peccei:1977hh,Peccei:1977ur,Wilczek:1977pj,Weinberg:1977ma} is an attractive extension of the Standard Model (SM) of particle physics, as it can solve the strong CP problem~\cite{Dine:2000cj,Strumia:2025ucr,Benabou:2025viy} while providing a compelling dark matter (DM)~candidate~\cite{Abbott:1982af,Dine:1982ah,Preskill:1982cy}. Inspired by the QCD axion, axion-like particles (ALPs) share properties with the QCD axion but cannot solve the strong CP problem. ALPs are theoretically well-motivated, appearing in many theories beyond the SM, and have a wide range of phenomenological signatures~\cite{Graham:2015ouw,Sikivie:2020zpn}.

Within four-dimensional effective field theory (EFT), the QCD axion and ALPs are typically pseudo-Goldstone bosons arising from a spontaneously-broken approximate global symmetry, called the Peccei-Quinn (PQ) symmetry. Their masses are set by anomalous or explicit sources of symmetry breaking, such as non-perturbative QCD instantons. As a result, they are vulnerable to quantum gravitational corrections that break global symmetries~\cite{Harlow:2018jwu,Alvey:2020nyh}. In the case of the QCD axion, gravitational corrections could reintroduce the strong CP problem by dominating QCD instanton effects. This is known as the axion quality problem~\cite{Georgi:1981pu,Barr:1992qq,Dine:1986bg,Kamionkowski:1992mf,Holman:1992us,Ghigna:1992iv}. In the more general case of ALPs, they could introduce a naturalness problem by dominating the tree-level ALP mass~\cite{Banerjee:2022wzk,Ghosh:2026ynw}. 

It is thus necessary to fine-tune ALP parameters or to provide a mechanism that delays the onset of gravitational corrections. For example, gauge symmetries could accidentally protect up to $K$ units of global charge~\cite{Randall:1992ut}. For example, local symmetries could accidentally protect PQ~\cite{Randall:1992ut,Dobrescu:1996jp,Redi:2016esr,Fukuda:2017ylt,DiLuzio:2017tjx,Ardu:2020qmo,Babu:2024qzb,DiLuzio:2025jhv} or the axion could be embedded into a higher-dimensional gauge field~\cite{Cheng:2001ys,Reece:2023czb}. In this work, we attempt to draw sharper predictions about the QCD axion and ALPs guided by quality and standard ideas about quantum gravitational corrections. We show that quality, measured statistically, complements phenomenological viability; masses and couplings allowed by phenomenology may require substantial fine-tuning. From this perspective, quality acts as a naturalness filter on the ALP parameter space.

\section{The Strong CP problem}\label{sec:strong_cp_problem}

The SM gauge groups permit a CP violating operator in the QCD sector,
\begin{equation}
    \lag \supset \thetaStrong\,\frac{\alpha_s}{8\pi}G^{a}_{\mu\nu}\tilde{G}^{\mu\nu}_a, 
    \label{eq:strong_CP_violation}
\end{equation}
where $\thetaStrong$ is an unknown parameter, $G$ and $\tilde{G}$ are the gluon field strength tensor and its dual, and $\alpha_s$ is the strong coupling constant. Considering quark redefinitions, the physically observable CP violating phase is
\begin{equation}\label{eq:theta_bar}
  \thetaTotal \equiv \thetaStrong - \thetaYukawa,
\end{equation}
where $\thetaYukawa$ originates from quark Yukawa
couplings. Measurements of the neutron electric dipole moment (nEDM) require that $|\thetaTotal| < 1.2 \times 10^{-10}$
at 90\% CL~\cite{Dragos:2019oxn,Abel:2020pzs}. Why do the CP violating phases in the strong and Yukawa sectors cancel so precisely in the SM? This fine-tuning is the strong CP problem~\cite{Dine:2015xga,Craig:2022eqo}.

The problem may be solved by promoting the phase to a dynamical field called the QCD axion. The QCD axion, and more generally ALPs, experience an effective low-energy potential,
\begin{equation}\label{eq:potential_ir}
V_\text{IR} = - m_\text{IR}^2 f_a^2 \cos\left(\frac{a}{f_a} + \thetaTotal\right),
\end{equation}
where $a$ is the axion field, $f_a$ its decay constant, $\thetaTotal$ a phase, and $m_\text{IR}$ the mass in the absence of UV corrections.\footnote{We have assumed a domain wall number $\Nanom = 1$, as possible in simple KSVZ models~\cite{Kim:1979if,Shifman:1979if} and more complex DSFZ models~\cite{Cox:2023squ}.} For the QCD axion,  the color anomaly generates a phase that exactly matches \cref{eq:theta_bar} and barrier height~\cite{DiVecchia:1980yfw} 
\begin{equation}
m_\text{IR}^2 f_a^2 \simeq f_\pi^2 m_\pi^2.    
\end{equation}
The minimum of the IR potential occurs at,
\begin{equation}
    \thetaEff \equiv \left\langle\frac{a}{f_a}\right\rangle + \thetaTotal = 0.
\end{equation}
For the QCD axion, this solves the strong CP problem.

\section{The quality problem}\label{sec:quality_problem}

Non-perturbative gravitational effects may result in additional effective non-renormalizable operators of the form~\cite{Kallosh:1995hi} 
\begin{equation}\label{eq:qg}
    \lag_\text{UV} =  \sum_{d=5}^\infty \sum_{k=1}^d c_{kd} \frac{|\Phi|^{d - k} \, \Phi^{k}}{\mpl^{d-4}} + \text{h.c.}
\end{equation}
where $\mpl$ is the reduced Planck mass, $\Phi$ is a complex scalar field, $c_{kd}$ are unknown complex Wilson coefficients, and $(d - k)$ must be a non-negative even number. These operators break the PQ symmetry by $k$ units of charge and are suppressed by dimension $d$. After spontaneous breaking of the PQ symmetry and integrating out the radial mode, the complex scalar can be parameterized as,
\begin{equation}
\Phi  = \frac{f_a}{\sqrt{2}} e^{i a / f_a},   
\end{equation}
which leads to a contribution to the potential,
\begin{equation}\label{eq:potential_uv}
V_\text{UV} = - 2 \mpl^4 \sum_{d=5}^\infty \sum_{k=1}^d \left|c_{kd}\right| \, \biggl(\frac{f_a}{\sqrt{2}\mpl}\biggr)^{d} \! \cos\biggl(\frac{k a}{f_a} + \phi_{kd}\biggr),
\end{equation}
where $c_{kd} = \left|c_{kd}\right| e^{i \phi_{kd}}$. From an EFT perspective, we consider forbidding operators in \cref{eq:qg} below a particular dimension, $D$, such that $c_{k5} = c_{k6} = \cdots = c_{k(D-1)} = 0$.

There are two effects of these UV contributions to the potential. First, they shift the minimum of the axion potential such that $\thetaEff \neq 0$. For the QCD axion, this spoils the solution to the strong CP problem. For example, considering only the $k = 1$ terms in the corrections in \cref{eq:potential_uv}, by treating the gravitational corrections as a perturbation, we find 
\begin{equation}\label{eq:theta_eff_approx}
    \thetaEff \simeq - \sum_{d=5}^\infty \frac{|c_{1d}| f_a^{d-2} \sin (\phi_{1d} - \thetaTotal)}{2^{d/2 -1}\mpl^{d-4} m_\text{IR}^2}. 
\end{equation}
Solving the strong CP problem thus requires one of the following: suppression of the Wilson coefficients, $|c_{1d}| \ll 1$; sufficient Planck suppression of the operators; close alignment of the phases, $|\phi_{1d} - \thetaTotal| \simeq 0$; or fine-tuning among the terms in the sum.

Second, the axion mass is the sum of IR and UV contributions,
\begin{align}
    m_a^2 =& \left\langle \frac{\partial^2 V_\text{IR}}{\partial a^2}\right\rangle + m_\text{UV}^2,
\end{align}
where we define
\begin{equation}
    m_\text{UV}^2 \equiv \left\langle \frac{\partial^2 V_\text{UV}}{\partial a^2}\right\rangle.
\end{equation}
Reminiscent of the hierarchy problem, this requires fine-tuning to maintain $m_a^2 \ll |m_\text{UV}^2|$. In general
\begin{equation}
    \frac{m_\text{UV}^2}{m_\text{IR}^2} = \sum_{d=5}^\infty\sum_{k=1}^d \frac{k^2 |c_{kd}|f_a^{d-2}\cos(k \thetaEff - k\thetaTotal + \phi_{kd})}{2^{d/2 - 1} \mpl^{d-4} m_\text{IR}^2}.
\end{equation}
Taking again the $k = 1$ terms and treating the gravitational corrections as a perturbation,
\begin{equation}\label{eq:m_uv_approx}
    \frac{m_\text{UV}^2}{m_\text{IR}^2} \simeq \sum_{d=5}^\infty \frac{|c_{1d}|f_a^{d-2}\cos(\phi_{1d} - \thetaTotal)}{2^{d/2 - 1} \mpl^{d-4} m_\text{IR}^2}. 
\end{equation}
Comparing \cref{eq:theta_eff_approx,eq:m_uv_approx} we see that suppressing the Wilson coefficients solves both $\theta_{\rm eff}\to 0$ and $|m_\text{UV}| \ll m_\text{IR}$ simultaneously, which is untrue for aligning phases or fine-tuning between terms.

\section{Quantifying fine-tuning}\label{sec:fine-tuning}

We quantify the quality problem by measuring the tuning required to mitigate the impact of PQ violating corrections to the axion potential. This depends on unknown Lagrangian parameters and Wilson coefficients. We marginalize the unknown Wilson coefficients over a prior, while treating the axion mass and decay constant as coordinates~\cite{Fowlie:2024dgj}. 

As a baseline, we assume $\mathcal{O}(1)$ magnitudes and agnosticism about phases of the Wilson coefficients. To construct this baseline scenario, we assume that the real and imaginary parts of the Wilson coefficients are independently and identically distributed as standard normal variables,
\begin{equation}\label{eq:gaussian_prior}
    \Im c_{kd} \sim \mathcal{N}(0, 1) \quad\text{and}\quad
    \Re c_{kd} \sim \mathcal{N}(0, 1).
\end{equation}
This corresponds to a uniform distribution on the phases and a Rayleigh distribution on the magnitudes,
\begin{equation}
    \phi_{kd} \sim \mathcal{U}(-\pi, \pi) \quad\text{and}\quad
    |c_{kd}| \sim \mathcal{R}(1),
\end{equation}
where $\mathcal{R}(1)$ denotes a Rayleigh distribution with scale parameter equal to one.  Increasing the variance in \cref{eq:gaussian_prior} would exacerbate the impact of gravitational corrections; decreasing it would imply a systematic suppression, which we do not wish to assume. 

For the QCD axion, we quantify the fine-tuning required to satisfy nEDM measurements using a Bayes factor~\cite{Jeffreys:1939xee,Kass:1995loi},
\begin{equation}\label{eq:bayes_factor}
    B = \frac{P(|\thetaEff| < 10^{-10} \given \text{QCD axion})}{P(|\thetaEff| < 10^{-10}  \given \text{SM-like model})},
\end{equation}
comparing an SM-like model, in which the strong CP phase is treated as an input, and a QCD axion model, in which the strong CP phase is computed from the minimum of the axion potential.

For a general ALP, we quantify fine-tuning of the IR mass parameter using a Bayes factor in favor of a model in which the ALP mass is an input parameter. For fixed Wilson coefficients, this exactly equals~\cite{Fowlie:2024nhs} the well-known Barbieri-Giudice (BG) measure~\cite{Ellis:1986yg,Barbieri:1987fn}:
\begin{equation}
\BG \equiv  
\left| \frac{m_\text{IR}^2}{m_a^2} \frac{\partial m_a^2}{\partial m_\text{IR}^2} \right| 
\end{equation}
We marginalize over the unknown Wilson coefficients, such that,
\begin{equation}\label{eq:alp_bf}
    \frac1B = \int \frac{1}{\BG}  \prod_{d=5}^\infty \prod_{k=1}^d  p(c_{kd}) \dint c_{kd}
\end{equation}
Because phases are marginalized, fine-tunings associated with alignment of phases are accounted for by these Bayes factors. The Jeffreys scale~\cite{Jeffreys:1939xee} for interpreting Bayes factors considers $B > 100$ to be decisive evidence. As we shall see, the Bayes factors associated with fine-tuning due to the quality problem dramatically exceed this.

\section{Quality control for the strong CP phase}\label{sec:strong_cp}

First, consider the strong CP problem in a toy version of the SM in which $\thetaEff$ is an input parameter. Treating $\thetaEff$ as an angle with a flat prior, we find
\begin{equation}\label{eq:toy}
    P(|\thetaEff| \le 10^{-10} \given \text{SM-like model}) 
    \simeq 10^{-10}.
\end{equation}
We now recognize the separate strong and Yukawa contributions to the CP phase in \cref{eq:theta_bar}. If we interpret $\thetaStrong$ and $\thetaYukawa$ as angles defined on a circle, as implied by the axial rotation, and if we consider $\thetaStrong$ and $\thetaYukawa$ to be uncorrelated, reflecting their origins in the strong and Yukawa CP sectors, the only permissible prior for $\thetaEff$ is again a flat prior. Thus, we obtain an identical result to that in \cref{eq:toy}. 

On the other hand, QCD axion models predict $\thetaEff = 0$
for any choices of parameters. Thus,
\begin{equation}
    P(|\thetaEff| \le 10^{-10} \given \text{QCD axion}) = 1.
\end{equation}
Considering only the data from nEDM measurements, a QCD axion model would be favored versus the SM by a Bayes factor
\begin{equation}
    B = \frac{P(|\thetaEff| \le 10^{-10} \given \text{QCD axion})}{P(|\thetaEff| \le 10^{-10} \given \text{SM-like model})} 
    \sim 10^{10}.
\end{equation}
However, this may be spoiled by gravitational corrections.

\begin{figure}[t]
    \centering
    \includegraphics[width=0.95\linewidth]{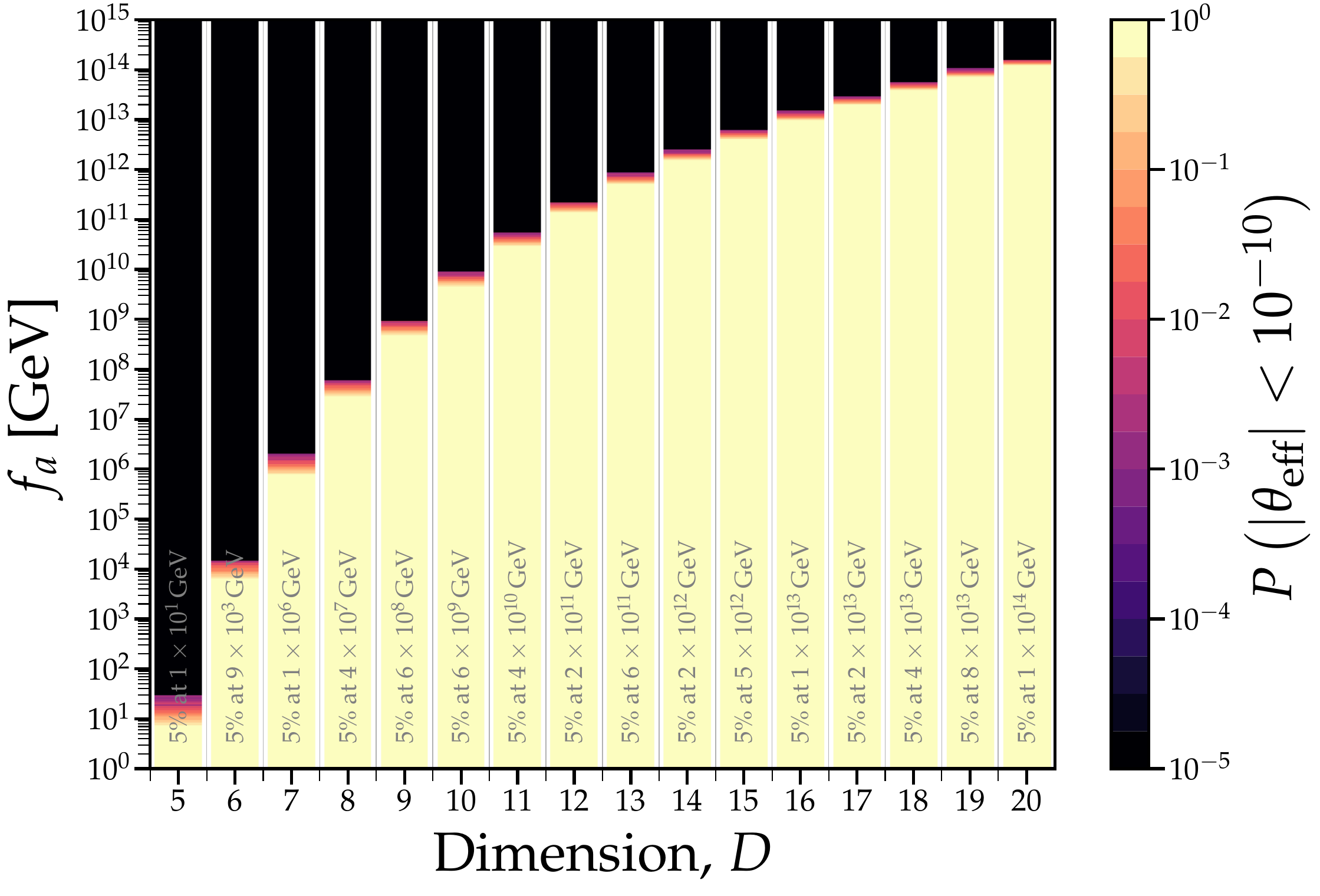}
    \caption{The probability of $|\thetaEff| < 10^{-10}$ as a function of $f_a$ and the dimension at which gravitational corrections first appear unsuppressed. The annotations show the value of $f_a$ required for $P(|\thetaEff| < 10^{-10}) = 5\%$.}
    \label{fig:dim}
\end{figure}

Rather than making assumptions about $f_a$ and the dimension at which gravitational corrections first appear, $D$, in \cref{fig:dim} we show the probability that $|\thetaEff| < 10^{-10}$ as a function of both $D$ and $f_a$, marginalizing over the Wilson coefficients. Once $f_a$ is large enough that gravitational corrections are relevant, the probability changes rapidly from one to zero, and the axion solution to the strong CP problem is ruined. We see that if we want that probability to be greater than $5\%$ and $f_a \gtrsim 10^{8}\gev$, we require $D \ge 9$.

Lastly, we consider the interplay of DM phenomenology and future experimental measurements of the nEDM. We construct three benchmarks for a QCD axion that constitutes all of the observed DM abundance:
\begin{enumerate}[label=\itshape(\roman*)]
\item \textit{Post-inflationary misalignment.}
Assuming the PQ symmetry breaks after inflation, the initial misalignment angle takes
independent values in each causal Hubble patch. The DM abundance depends on the anharmonic average of the initial misalignment, $\theta_0 \simeq 2.15$~\cite{GrillidiCortona:2015jxo}.\footnote{We neglect the impact of gravitational corrections on the anharmonic average angle.} 
Combined with the observed DM abundance, this fixes $f_a \simeq 2\times10^{11}\gev$ and thus $m_a \simeq 30\muev$. To satisfy existing nEDM constraints, this requires $D \ge 13$.

\item \textit{Post-inflationary cosmic-string production.}
In the same post-inflationary scenario, a network of global cosmic strings forms and
radiates axions with a momentum spectrum $F(k)\propto k^{-q}$.
For a spectral index $q>1$ (IR-dominated emission), the string contribution to the
DM abundance dominates the misalignment contribution by about two orders of magnitude~\cite{Chathirathas:2025aan}. The correct abundance thus requires a smaller decay constant, $f_a \simeq 1.60\times10^{10}\gev$, and a heavier mass, $m_a \simeq 3.47\times10^{-4}\ev$. Existing nEDM constraints now require $D \ge 11$.

\item \textit{Pre-inflationary misalignment.}
Assuming the PQ symmetry breaks before inflation, the initial misalignment angle is a single, spatially uniform value
that is not averaged, and that sets the DM abundance. To create a scenario that is distinct from post-inflationary misalignment, we seek $\theta_0 \ll 2.15$. To avoid excessive fine-tuning of a tiny angle, we take $\theta_0 = \pi/100$. This yields $f_a \simeq 6\times10^{12}\gev$, greater than both post-inflationary cases, and $m_a \simeq 9\times10^{-7}\ev$. Existing nEDM constraints now require $D \ge 16$.
\end{enumerate}

\begin{figure}[t]
    \centering
    \hspace{-3mm}
    \includegraphics[width=0.95\linewidth]{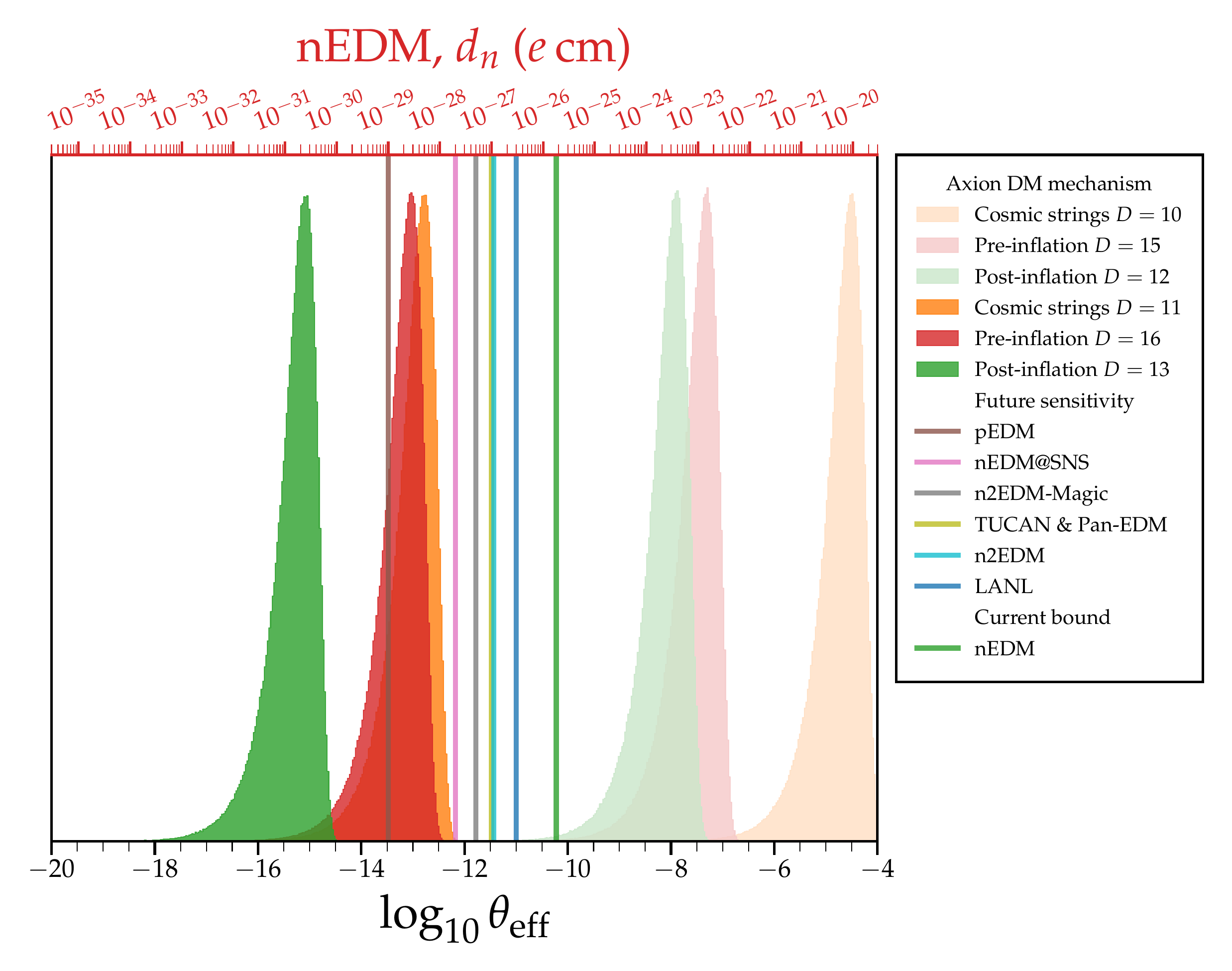}
    \caption{Distribution of $\thetaEff$ when the QCD axion plays the role of DM, produced by cosmic strings, and pre- and post-inflationary alignment, and gravitational corrections first appear at dimension $D$. Present and projected limits on the nEDM are shown.}
    \label{fig:dm_nedm_interplay}
\end{figure}

We show the residual strong CP phase predicted in these scenarios 
in \cref{fig:dm_nedm_interplay} as well as the current bound, 
$180\times 10^{-28}\ecm$ 
at $90\%$ CL~\cite{Abel:2020pzs}, and the projected improved bounds from several future nEDM experiments~\cite{Alarcon:2022ero}, which are between $2$ -- $30\times 10^{-28}\ecm$~\cite{Imam:2020fvu,Lauss:2025n2EDM,Wurm:2019yfj,TUCAN:2025rxj,nEDM:2022nqd,Lauss:2025n2EDM,Ito:2017ywc}.
Finally, future measurements of the proton EDM (pEDM) with sensitivity of around $10^{-29}\ecm$ may translate into similar constraints on the nEDM~\cite{pEDM:2022ytu}. The QCD axions in these three scenarios satisfy current bounds on the nEDM, provided that gravitational corrections first appear at sufficiently high operator dimension. However, this takes them beyond most future nEDM experiments. The cosmic strings and post-inflationary alignment scenarios are in the projected experimental reach of the future pEDM experiment.

\section{Quality control for ALP mass}\label{sec:alps}

ALP models are typically treated as effective theories, parameterized by the physical mass of the ALP, $m_a$, and the axion decay constant, $f_a$. Since ALP models do not attempt to solve the strong CP problem, only fine-tuning of the ALP mass is relevant. ALPs are usually considered extremely light, $m_a \ll 1\gev$, though there are exceptions~\cite{Murayama:2026ioh,Zamoro:2026ily}. We present fine-tuning maps using \cref{eq:alp_bf} on the ($m_a$, $1/f_a$) plane of ALP parameter space by averaging over 100,000 Monte Carlo realizations of Wilson coefficients.

\begin{figure}[t]
    \centering
    \includegraphics[width=0.95\linewidth]{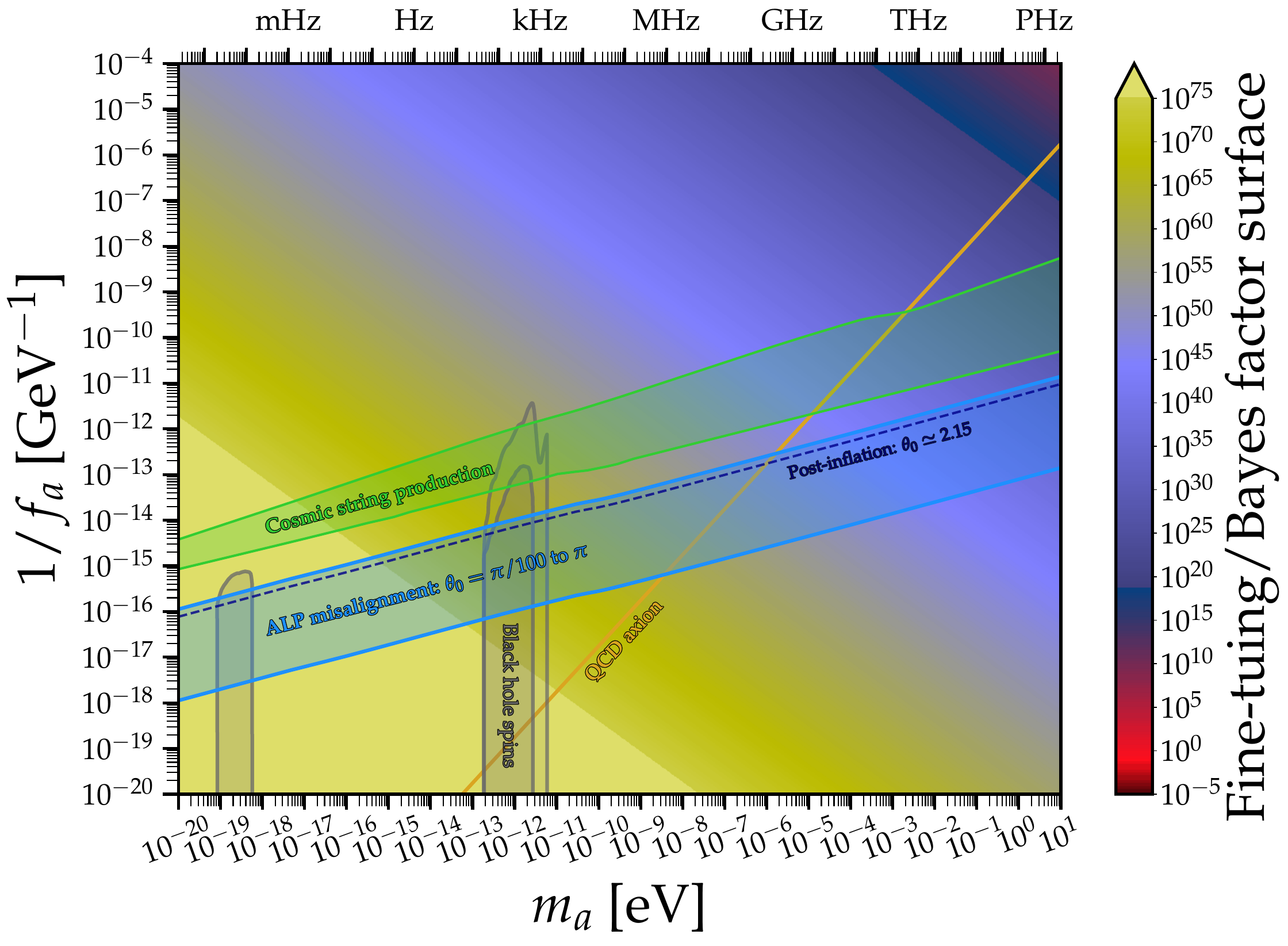}
    \caption{The fine-tuning price on the ($m_a$, $1/f_a$) plane, assuming that gravitational corrections appear at $d = 5$. We show regions that produce the correct DM abundance using cosmic string production (green region) with the $q>1$ results from ref.~\cite{Chathirathas:2025aan} and allowing for variation in $m_a(T)$; and misalignment (blue region), assuming standard cosmology and $\pi/100 \le \theta_0 \le \pi$~\cite{Hui:2016ltb,Hook:2018dlk}. We show the black hole superradiant constraints (gray regions)~\cite{Hoof:2024quk,Witte:2024drg}, which do not assume any axion-SM couplings other than gravity. Finally, for context we show the QCD axion (gold line)~\cite{OHare:2024nmr}.}
    \label{fig:alp_fine_tuning_map_5}
\end{figure}

\begin{figure}[t]
    \centering
    \includegraphics[width=0.95\linewidth]{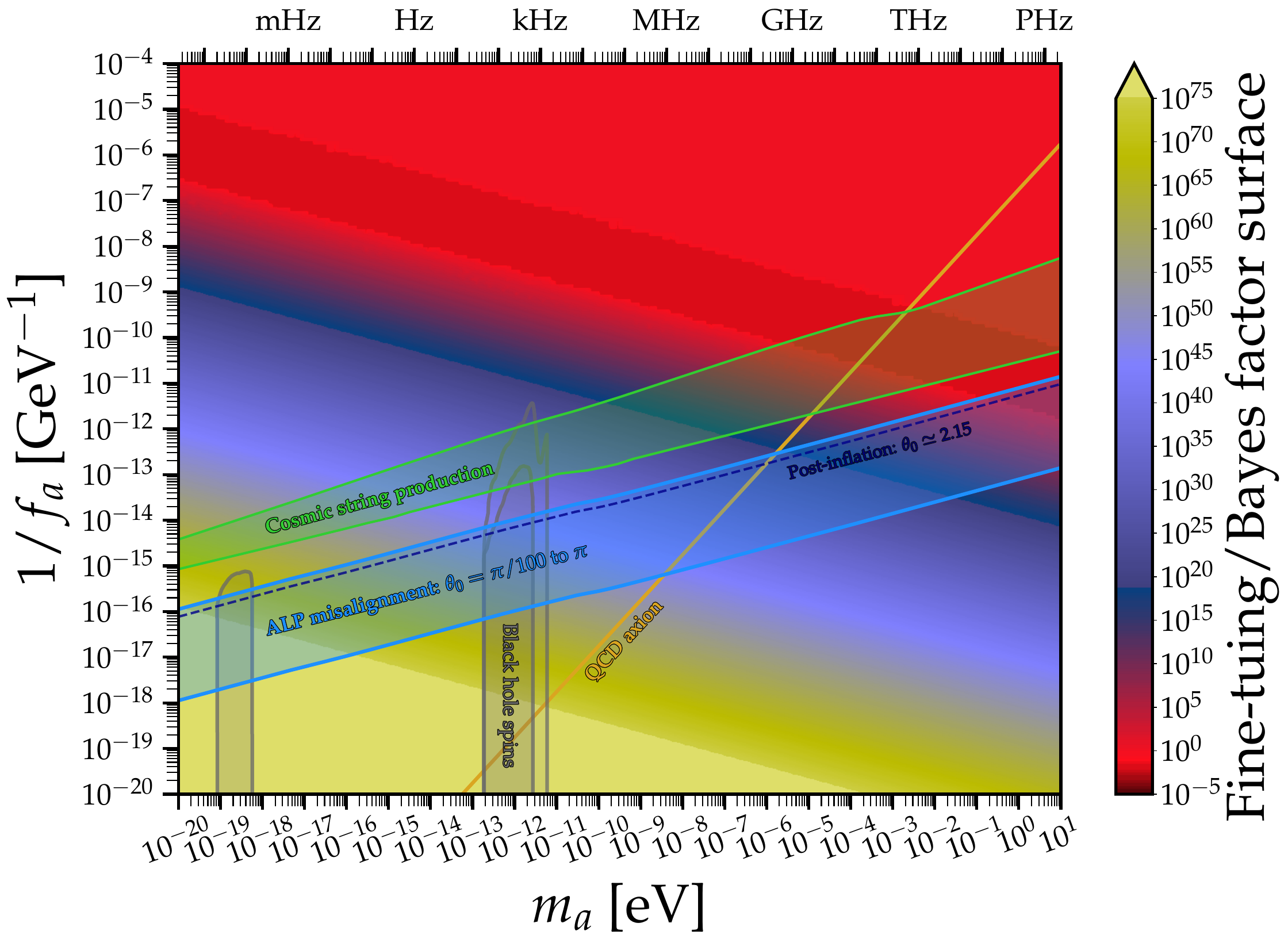}
    \caption{The fine-tuning price on the ($m_a$, $1/f_a$) plane, assuming that gravitational corrections appear at $d = 10$. The annotations are identical to those in \cref{fig:alp_fine_tuning_map_5}.}
    \label{fig:alp_fine_tuning_map_10}
\end{figure}

\Cref{fig:alp_fine_tuning_map_5} shows fine-tuning on the ALP ($m_a$, $1/f_a$) plane, assuming that gravitational corrections start at $d = 5$. The fine-tuning penalty is complementary to experimental constraints. We observe that smaller masses and larger $f_a$ are severely fine-tuned to more than $10^{75}$ and we require $m_a \simeq 10\ev$ and $f_a \sim 10^4\gev$ to avoid tuning. We show the QCD axion relationship between mass and decay constant, though note that, due to \cref{eq:theta_eff_approx,eq:m_uv_approx}, the tuning associated with the strong CP problem is more severe, and that both tuning problems may be solved at the same time. We relax the assumption that gravitational corrections start at $d=5$ in \cref{fig:alp_fine_tuning_map_10}, where we assume that they are absent until $d=10$. Although fine-tuning remains severe for lighter ALP mass and larger $f_a$, there is a band of parameter space at $m_a \gtrsim 1\mev$ and $f_a \lesssim 10^{12}\gev$ that is not tuned.

Neglecting coefficients in \cref{eq:m_uv_approx}, for $m_a^2 \gtrsim m_\text{UV}^2$ we approximately require
\begin{equation}\label{eq:f_a_approx}
    \log f_a \lesssim \frac{2}{D-2} \log m_a + \frac{D-4}{D-2} \log \mpl.
\end{equation}
Thus, the contours in \cref{fig:alp_fine_tuning_map_5,fig:alp_fine_tuning_map_10} are at gradient $-2/(D-2)$ and, intuitively, even as $D\to\infty$ we require $f_a \lesssim \mpl$, corresponding to the limit of EFT validity. 

\begin{figure}[t]
    \centering
    \includegraphics[width=0.95\linewidth]{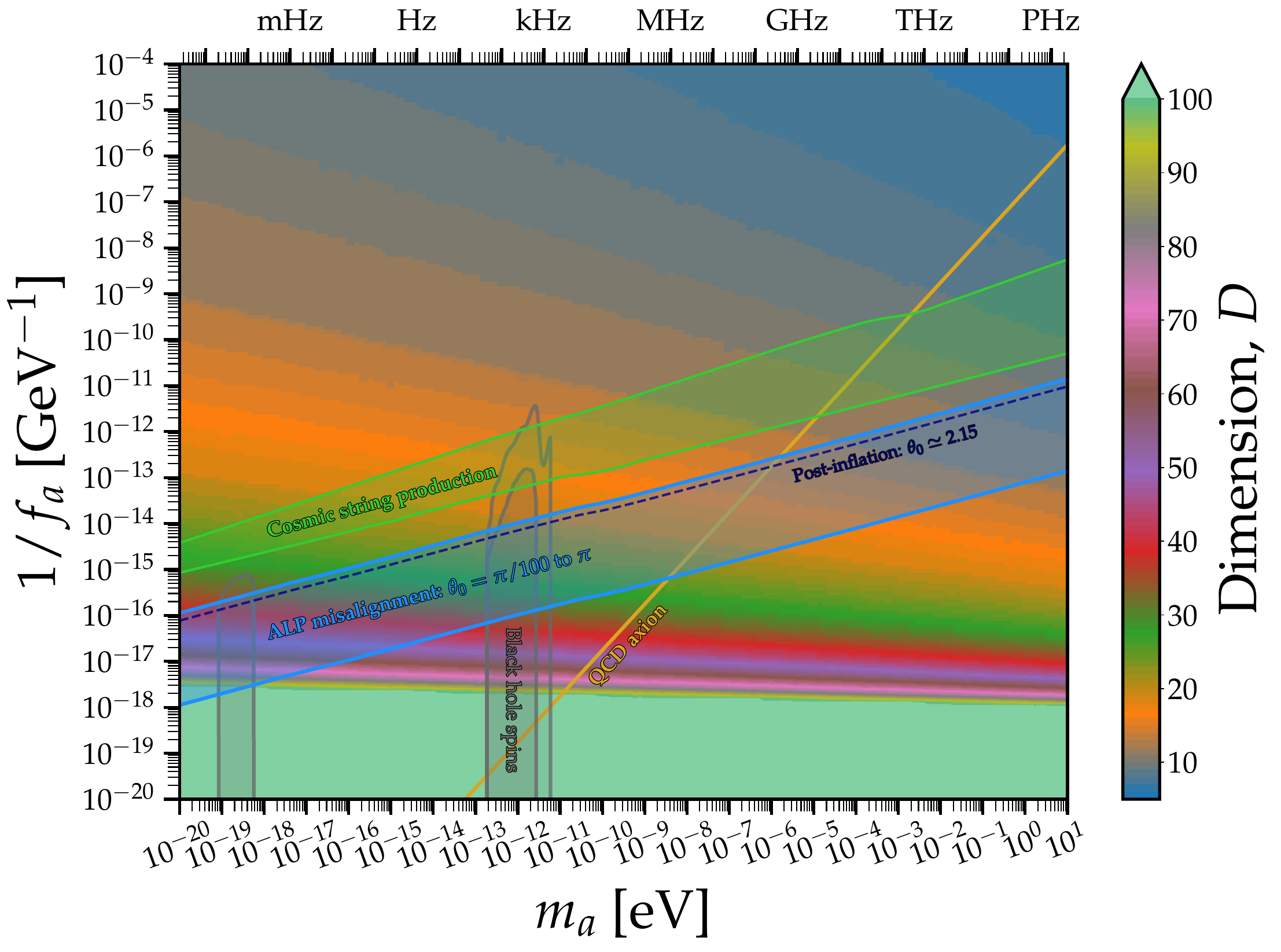}
    \caption{The dimension $D$ across the ($m_a$, $1/f_a$) plane below which the quality problem must be solved to avoid fine-tuning of 100 or more. The annotations are identical to those in \cref{fig:alp_fine_tuning_map_5}.}
    \label{fig:alp_dim}
\end{figure}

Lastly, in \cref{fig:alp_dim}, we show the dimension $D$ below which gravitational corrections must vanish if one is to obtain a fine-tuning of $B < 100$. This is important for constructing realistic UV completions of axion models. We see that $f_a \gtrsim 10^{16}\gev$ would require PQ protected up to about $D = 100$.  There is, however, parameter space that could accommodate $D \lesssim 10$. The space at $f_a \gtrsim \mpl$ and $m_a \ll \mpl$ cannot be untuned at any dimension.

There may be difficulties in forbidding operators up to dimension 100 using an accidental symmetry and thus it may be preferable to embed the axion into a higher-dimensional gauge group.

\section{Conclusions}\label{sec:concs}

Axion phenomenology is usually organized in terms of masses, decay constants, couplings, and experimental reach. We have argued that this parameter space should be supplemented by the degree of fine-tuning that every possible axion configuration requires.  If quantum gravity violates global symmetries, Planck-suppressed PQ-breaking operators generically perturb the axion potential. A point in parameter space is therefore not characterized only by whether it is allowed or discoverable, but also by how much UV protection is required for it to be natural.

We quantified this question by treating the quality problem as a fine-tuning problem. As a baseline, we assumed unsuppressed $\mathcal{O}(1)$ Wilson coefficients and random phases for the PQ-breaking operators, and marginalized over these unknown coefficients. This setup does not apply to UV completions that suppress PQ violation, align phases, or enforce cancellations; rather, it provides the reference expectation in the absence of such mechanisms.

For the QCD axion, Planck-suppressed corrections can shift the axion minimum away from the CP-conserving point and reintroduce the strong CP problem. For axion DM benchmarks, satisfying current nEDM bounds requires PQ-breaking operators to be absent or suppressed up to high dimensions, typically $D \gtrsim 11$--$16$, depending on the cosmological history. If the leading operators appear near the lowest dimensions allowed today, the residual strong CP phase may be within reach of future neutron or proton EDM searches. These experiments therefore probe not only CP violation, but also the quality of the axion solution.

For ALPs, the same corrections appear as a mass naturalness problem. Large regions of the usual ($m_a$, $1/f_a$) plane are highly tuned in the baseline scenario, or equivalently require PQ protection to high operator dimension. In particular, regions with large $f_a$ can require protection far beyond $D\sim 10$, making the required UV structure a central part of the model.
Our main conclusion is that axion parameter space is not theoretically uniform. Experimental searches tell us where axions and ALPs may be found; quality tells us where they are natural under standard expectations about quantum gravity. Regions that fail this baseline quality test are not excluded, but they demand explicit UV mechanisms that suppress PQ-breaking operators or align their phases.

\acknowledgments

\bigskip\noindent AF was supported by the National Natural Science Foundation of China (NNSFC) RFIS-II W2432006. The work of GH is supported by the Neutrino Theory Network Fellowship with contract number 726844. AC acknowledges the support of S. Ge, funded by the NSFC (Grant Nos. 12425506, 12375101, 12090060, and 12090064).

\appendix

\section{Flat distribution on angle}\label{app:flat}

To see this, consider the prior predictive distribution for $\thetaTotal$,
\begin{equation}
    p(\thetaTotal) = \int p(\theta) \, p(\thetaYukawa) \, \delta(\thetaTotal - (\theta - \thetaYukawa)) \dint \theta \dint \thetaYukawa.
\end{equation}
Integrating over a flat distribution for $\theta$, we obtain
\begin{equation}
    p(\thetaTotal) = \int p(\theta = \thetaTotal + \thetaYukawa) \, p(\thetaYukawa) \dint \thetaYukawa = \text{const.}  
\end{equation}
This was the convolution of the flat prior for $\theta$ and the prior for $\thetaYukawa$. The convolution of anything with a flat distribution gives a flat distribution. Thus, just as in \cref{eq:toy}, we find,
\begin{equation}
    P(|\thetaEff| \le 10^{-10} \given \text{SM-like model}) \simeq 10^{-10}
\end{equation}
Another way of thinking about this is that the parameter $\thetaYukawa$ means that the parameter $\theta$ must be fine-tuned about $\thetaYukawa$ rather than about $0$. Since the problem is shift invariant in $\theta$, this tuning is the same. 

\section{Analytic quality bound from the leading PQ-breaking operator}

It is useful to isolate an analytic lower bound on the operator dimension
at which PQ-violating corrections may first appear. We write the leading
dimension-$D$ correction to the axion potential as
\begin{equation}
V_{\rm UV}^{(D)}
=
- 2 |c_D| M_{\rm Pl}^4
\left(
\frac{f_a}{\sqrt{2}M_{\rm Pl}}
\right)^D
\cos\left(k\frac{a}{f_a}+\phi_D\right),
\end{equation}
where $k$ is the PQ charge of the operator. In this section we do not
assume a probability distribution for the Wilson coefficient. Instead, we
take a benchmark natural size
\begin{equation}
|c_D| = c_\star ,
\qquad
c_\star = 0.1 ,
\end{equation}
corresponding to a moderately suppressed but still order-one coefficient.
This should be viewed as an optimistic benchmark relative to unsuppressed
Planck-scale PQ violation, for which $|c_D|\sim 1$. We do not attach this
choice to a perturbative loop interpretation, since the gravitational
origin of the PQ-breaking operators need not be perturbative.

For the QCD axion, the infrared potential has curvature
\begin{equation}
\chi_{\rm QCD} \simeq m_a^2 f_a^2 .
\end{equation}
The leading UV correction shifts the minimum by an amount of order
\begin{equation}
|\delta\theta_{\rm eff}|
\sim
\frac{
2 k |c_D| M_{\rm Pl}^4
\left(f_a/\sqrt{2}M_{\rm Pl}\right)^D
}{
\chi_{\rm QCD}
},
\end{equation}
up to an order-one phase factor. Requiring this shift to be smaller than
the neutron-EDM bound, $|\theta_{\rm eff}|<\theta_{\max}$, gives the
analytic quality condition
\begin{equation}
2 k c_\star M_{\rm Pl}^4
\left(
\frac{f_a}{\sqrt{2}M_{\rm Pl}}
\right)^D
<
\theta_{\max}\chi_{\rm QCD}.
\end{equation}
For $f_a<\sqrt{2}M_{\rm Pl}$, this implies
\begin{equation}
D
>
\frac{
\log\!\left[
\dfrac{2 k c_\star M_{\rm Pl}^4}
{\theta_{\max}\chi_{\rm QCD}}
\right]
}{
\log\!\left[
\dfrac{\sqrt{2}M_{\rm Pl}}{f_a}
\right]
}.
\end{equation}
Equivalently, the minimum protection dimension for the QCD axion is
\begin{equation}
D_{\min}^{\rm QCD}(f_a)
=
\max\left\{
5,\,
\left\lceil
\frac{
\log\!\left[
\dfrac{2 k c_\star M_{\rm Pl}^4}
{\theta_{\max}\chi_{\rm QCD}}
\right]
}{
\log\!\left[
\dfrac{\sqrt{2}M_{\rm Pl}}{f_a}
\right]
}
\right\rceil
\right\}.
\end{equation}
This is the minimum dimension below which PQ-violating operators must be
absent, assuming no small phases or cancellations.

For a generic ALP, the same operator induces a UV contribution to the
mass,
\begin{equation}
|m_{\rm UV}^2|
\sim
\frac{
2 k^2 |c_D| M_{\rm Pl}^4
\left(f_a/\sqrt{2}M_{\rm Pl}\right)^D
}{
f_a^2
},
\end{equation}
again up to an order-one phase factor. Demanding that this correction not
exceed the physical ALP mass gives
\begin{equation}
\frac{
2 k^2 c_\star M_{\rm Pl}^4
\left(f_a/\sqrt{2}M_{\rm Pl}\right)^D
}{
f_a^2
}
<
m_a^2 .
\end{equation}
Solving for $D$ gives
\begin{equation}
D
>
\frac{
\log\!\left[
\dfrac{2 k^2 c_\star M_{\rm Pl}^4}
{m_a^2 f_a^2}
\right]
}{
\log\!\left[
\dfrac{\sqrt{2}M_{\rm Pl}}{f_a}
\right]
}.
\end{equation}
The corresponding minimum protection dimension is therefore
\begin{equation}
D_{\min}^{\rm ALP}(m_a,f_a)
=
\max\left\{
5,\,
\left\lceil
\frac{
\log\!\left[
\dfrac{2 k^2 c_\star M_{\rm Pl}^4}
{m_a^2 f_a^2}
\right]
}{
\log\!\left[
\dfrac{\sqrt{2}M_{\rm Pl}}{f_a}
\right]
}
\right\rceil
\right\}.
\end{equation}
This expression defines an analytic quality bound across the ALP
$(m_a,f_a)$ plane. For $k=1$ it gives the most optimistic leading-harmonic
estimate; larger $k$ strengthens the bound only logarithmically.

\bibliographystyle{apsrev4-2j}
\bibliography{refs}    

\end{document}